\documentclass[12pt,twoside]{article}
\usepackage{fleqn,espcrc1}

\usepackage{graphicx}
\usepackage[figuresright]{rotating}

\newcommand{\AmS}{{\protect\the\textfont2
  A\kern-.1667em\lower.5ex\hbox{M}\kern-.125emS}}

\newcommand{\beq}{\begin{equation}}
\newcommand{\eeq}{\end{equation}}
\newcommand{\bce}{\begin{center}}
\newcommand{\ece}{\end{center}}
\newcommand{\bea}{\begin{eqnarray}}
\newcommand{\eea}{\end{eqnarray}}

\newcommand{\ave}[1]{\langle {#1} \rangle}
\newcommand{\tave}[1]{\langle\!\langle{#1}\rangle\!\rangle}

\title{Meson Properties in Dense Hadronic Matter}

\author{J. Wambach\address{ Institut f{\"u}r Kernphysik, 
        Technische Universit{\"a}t Darmstadt,\\ 
        Schlo{\ss}gartenstr. 9, 
        D-64289 Darmstadt, Germany}%
        \thanks{This work was supported by GSI and BMBF},
        }
       
\begin{document}

\maketitle

\begin{abstract}
The medium modification of mesons in dense hadronic matter is discussed
with a focus on the relationship to the chiral structure of the non-perturbative
QCD vacuum. 
\end{abstract}

\section{INTRODUCTION}

One of the fundamental questions in strong interaction physics concerns the
generation of mass in the sector of light quarks. 
How can nucleons with
a mass of roughly 1 GeV arise from up and down quark masses of less than 10 MeV? 
The answer lies in the non-perturbative structure of the QCD vacuum itself in 
which quarks and gluons condense. This is in marked contrast to the heavy-quark 
sector where the masses of the hadrons are determined by the quark masses 
themselves. These, in turn, are set at the electroweak scale through the Higgs 
mechanism. 

When nuclear matter is subjected to extreme conditions in density and temperature
such as in the interior of neutron stars or in central relativistic heavy-ion
collisions, the QCD vacuum will be altered, eventually leading 
to the liberation of the elementary constituents in a new state of matter, the 
'quark gluon plasma' (QGP). Such a restructuring of the vacuum must be accompanied by 
significant changes in the spectral properties of hadrons, commonly called 
'medium modifications'. Here mesons are of particular importance, since they
constitute the 'elementary' $\bar qq$ excitations of the vacuum and hence serve
as experimental probes of the underlying vacuum changes. For the structure
of light mesons chiral symmetry and its spontaneous break-down plays a decisive 
role as is well established from lattice QCD simulations~\cite{Nege}. The
physical mechanism for spontaneous symmetry breaking and mass generation is 
provided by instantons which largely saturate the euclidian gauge-field 
action~\cite{SchaSh}. These induce effective quark-(anti)quark 
interactions which are strong enough to cause a BCS-like transition~\cite{Nambu}
to a $\bar qq$ condensate. As a result a mass gap appears, the 'constituent quark'
mass of $\sim$ 300-500 MeV. It has been demonstrated that this picture yields an 
excellent description of hadronic correlation functions~\cite{SchaSh} and
hence forms the basis for a variety of approaches to deal with
medium modifications. An obvious starting point is NJL-type models~\cite{Nambu} 
where the instanton-induced interaction is assumed to be point-like.
They are very successful in
describing mesonic spectral functions, especially  when treated beyond the 
mean-field approximation~\cite{OeBuWa}. Such calculations are very
demanding but finally lead to results in accordance with 
'effective theories' (such as the vector dominance model (VDM)~\cite{Saku}) 
in which hadronic fields feature as the elementary degrees of freedom. It
is more 'economical' to start with those from the outset, especially when dealing
with 'precursor' phenomena of the QGP transition. This strategy will be
followed in the present contribution. 

\section{CHIRAL EFFECTIVE THEORY AND CORRELATORS}
The spontaneous breaking of chiral symmetry 
has two important consequences. One is the appearance of massless Goldstone 
bosons and the other is the absence of 
parity doublets in the hadron spectrum ($m_\pi\neq m_{f_0}, m_\rho\neq m_{a_1}$ etc).
For the present discussion the second will be more relevant. 
Chiral symmetry does, however, more than just predict the existence of Goldstone
bosons. It also prescribes and severely restricts their mutual
interactions as well as those with other hadrons. This forms the basis for
constructing 'chiral effective theories'.

The formal starting point are the pertinent quark currents
\beq
J_i(x)=\bar q(x)\Gamma_i q(x), \Gamma_i=1, \gamma_5,\gamma_\mu,..
\eeq
which are identified with {\it elementary} hadronic fields $\phi_i(x)$.
One then writes down the most general effective Lagrangian, consistent with 
the underlying symmetries and anomaly structure of QCD. Relevant examples are
the linear- or nonlinear sigma model, gauged sigma models, etc. 

The physical information about hadronic spectral properties and their
in-medium modi-
fication is contained in the current-current 
correlation functions $D_i(x)\propto\ave{J_i(x)J_i(0)}$. In terms of the elementary 
hadronic fields and for matter in thermal and chemical equilibrium they are 
given by the (retarded) momentum-space Green's function
\beq
D_i(\omega,\vec q)=i\!\int\!\!d^4x\,e^{iqx}\theta(x_0)
\tave{[\phi_i(x),\phi_i(0)]}
\eeq
where the average is taken in the grand canonical ensemble. All information 
about the physical excitations of the medium is contained in the spectral function
\beq							   
A_i(\omega,\vec q)=
-{1\over \pi}{\rm Im}D_i(\omega,\vec q)\, .
\eeq
In the first step the parameters of the chiral Lagrangian are adjusted to
reproduce the vacuum properties of the hadronic correlators in question
(mass, width..) as well as possible. Here either 
perturbative, or in some cases, non-perturbative methods~\cite{Aoui1} are used.
Once the vacuum model is fixed, the medium modifications can be predicted 
with a good degree of accuracy.

\section{MESONS IN THE HADRONIC MEDIUM}

\subsection{Chiral Condensate evolution}

With
increasing density (baryo-chemical potential $\mu_q$) and temperature $T$ 
the quark 
condensate will diminish and eventually vanish, thus restoring chiral symmetry.
For hadronic matter in equilibrium the QCD partition function is given by
\beq
{\cal Z}_{QCD}(V,T,\mu_q)={\rm Tr}e^{-(H_{QCD}-\mu_q N_q)/T},
\eeq
where $N_q$ is the quark
number operator and $\mu_q$ the quark chemical potential. The quark condensate
$\tave{\bar qq}$ can be directly inferred from the free energy density
in the thermodynamic limit
\beq
\Omega_{QCD}(T,\mu_q)=-\lim_{V\to\infty}{T\over V}\ln{\cal Z}_{QCD}(V,T,\mu_q)
\quad {\rm as}\quad 
\tave{\bar qq}={\partial\Omega_{QCD}(T,\mu_q)\over\partial m^\circ_q}~,
\eeq
where $m^\circ_q$ denotes the current quark mass. 

In the broken phase, the obvious first
step is to approximate the free energy density by an ideal gas of hadrons.
The resulting in-medium condensate then becomes
\beq
{\tave{\bar qq}\over\ave{\bar qq}}=1-
\sum_h{\Sigma_h\varrho^s_h(T,\mu_q)\over f_\pi^2 m_\pi^2};\quad 
\Sigma_h=m^\circ_q{\partial m_h\over \partial m^\circ_h}
\label{qqt}
\eeq
where $\varrho^s_h$ denotes  the scalar density of hadrons and $m_h$ their 
vacuum mass. At low temperature and small $\mu_q$ in which 
the hadron gas is
dominated by thermally excited pions and a free Fermi gas of
nucleons,  Eq.~(\ref{qqt}) leads to the model-independent leading-order result  
\beq
{\tave{\bar qq}\over\ave{\bar qq}}=1-{T^2\over 8f_\pi^2}
-0.3{\rho\over\rho_0}\dots~,
\label{lowden}
\eeq
where $\rho_0=0.16$/fm$^3$ is the saturation density of 
symmetric nuclear matter. Thus the mere presence
of an ideal gas of hadrons already alters the vacuum and
leads to a decrease of the condensate, without changing the vacuum
properties of the hadrons! Obviously medium-modifications 
and the corresponding non-trivial change of the QCD vacuum have to
involve hadronic interactions. They become increasingly important as
the matter grows hotter and denser, i.e. as the point
of chiral restoration is approached. The theoretical description
involves more and more degrees of freedom,
which severely limits the description in terms of elementary hadronic
fields near the phase transition.

\subsection{Fluctuations of the chiral condensate}

Being dependent on the renormalization scale, the chiral condensate is
not observable. However, the fluctuations 
$\tave{(\bar qq)^2}-\tave{\bar qq}^2$
are. They correspond to the scalar (chiral) susceptibility and are
given as the second derivative of the free energy density 
\beq
\chi_+=\tave{(\bar qq)^2}-\tave{\bar qq}^2=
{\partial^2\Omega_{QCD}(T,\mu_q)\over\partial^2 m^\circ_q}
\eeq
w.r.t. to the current quark mass.
To make contact with the hadronic spectrum we consider the bilinear
scalar quark current $J_+=\bar qq$. From the corresponding
correlator
\beq
D_+(\omega,\vec q)=i\int\!\!d^4x\,\theta(x_0)e^{iqx}
\tave{[(J_+(x),J_+(0)]}\,.
\label{scorr}
\eeq
the scalar susceptibility is obtained as the 'polarizability sum rule'
for the scalar spectral function $A_+(\omega,\vec q)$:
\beq
\chi_+=\int_0^\infty\!d\omega A_+(\omega,\vec q=0)
\label{chi}~, 
\eeq
and hence relates to the properties of the scalar $f_0$ meson  in the medium.
 
From direct lattice simulations it is established that
QCD with two flavors and vanishing $\mu_q$ exhibits a second-order phase 
transition in the chiral limit~\cite{Karsch1}. The transition at $T=0$
but finite $\mu_q$ is not as well understood, since it cannot be
simulated on the lattice, due to complex fermion determinants. Model
calculations indicate however that it may be first order at several
times $\rho_0$.
As a second- or weak first-order phase transition is approached the
fluctuations of the order parameter become large. A measure for the growth
are the appropriate susceptibilities which, in a second-order transition,
diverge with critical exponents that are determined by universal behavior.
The approach to criticality  in the case of two-flavor QCD
is clearly seen on the lattice~\cite{Karsch2}.

\begin{figure}[htb]
\bce
\includegraphics[width=15pc]{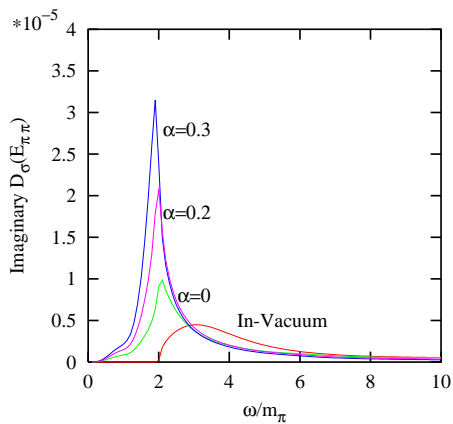}
\caption{The imaginary part of $D_\sigma(\omega,0)$~\cite{Schuck2} at
nuclear saturation density for various
values of $\alpha$, the parameter that controls the linear density
dependence of $\tave{\sigma}$ (Eq.~\ref{lowden}).}
\label{fig:sigpropf}
\ece
\end{figure}

Following the suggestion in ref.~\cite{PiWi} that two-flavor
QCD lies in the same universality class as the $O(4)$ Heisenberg model
we can use the linear sigma model to asses the chiral fluctuations in
the hadronic medium. In this case the scalar correlator $D_+$ in
Eq.~(\ref{scorr}) is identified with the in-medium $\sigma$-propagator:
\beq
D_+=D_\sigma(\omega,\vec q)=i\int\!\!d^4xe^{iqx}\theta(x_0)
\tave{[\sigma(x),\sigma(0)]}\, .
\eeq
Applying a low-density expansion for the in-medium condensate according
to Eq.~(\ref{lowden}) 
\beq
\tave{\sigma}\equiv\ave{\sigma}\Phi(\rho);\quad 
\Phi(\rho)=1-\alpha{\rho\over\rho_0}
\label{lowden1}
\eeq
together with pionic loop corrections
the results~\cite{Schuck2} depicted in Fig.~\ref{fig:sigpropf}
predict a dramatic near-threshold enhancement of the $\sigma$ spectral function
already at $\rho_0$ (corresponding to $\Phi(\rho)\sim 0.7-0.8$). It should be
noted that this effect is also found in the non-linear realization of chiral 
symmetry~\cite{Hats1} and is therefore generic.

\begin{figure}[htb]
\begin{minipage}[t]{80mm}
\hspace{0.75cm}\includegraphics[angle=90,width=11pc]{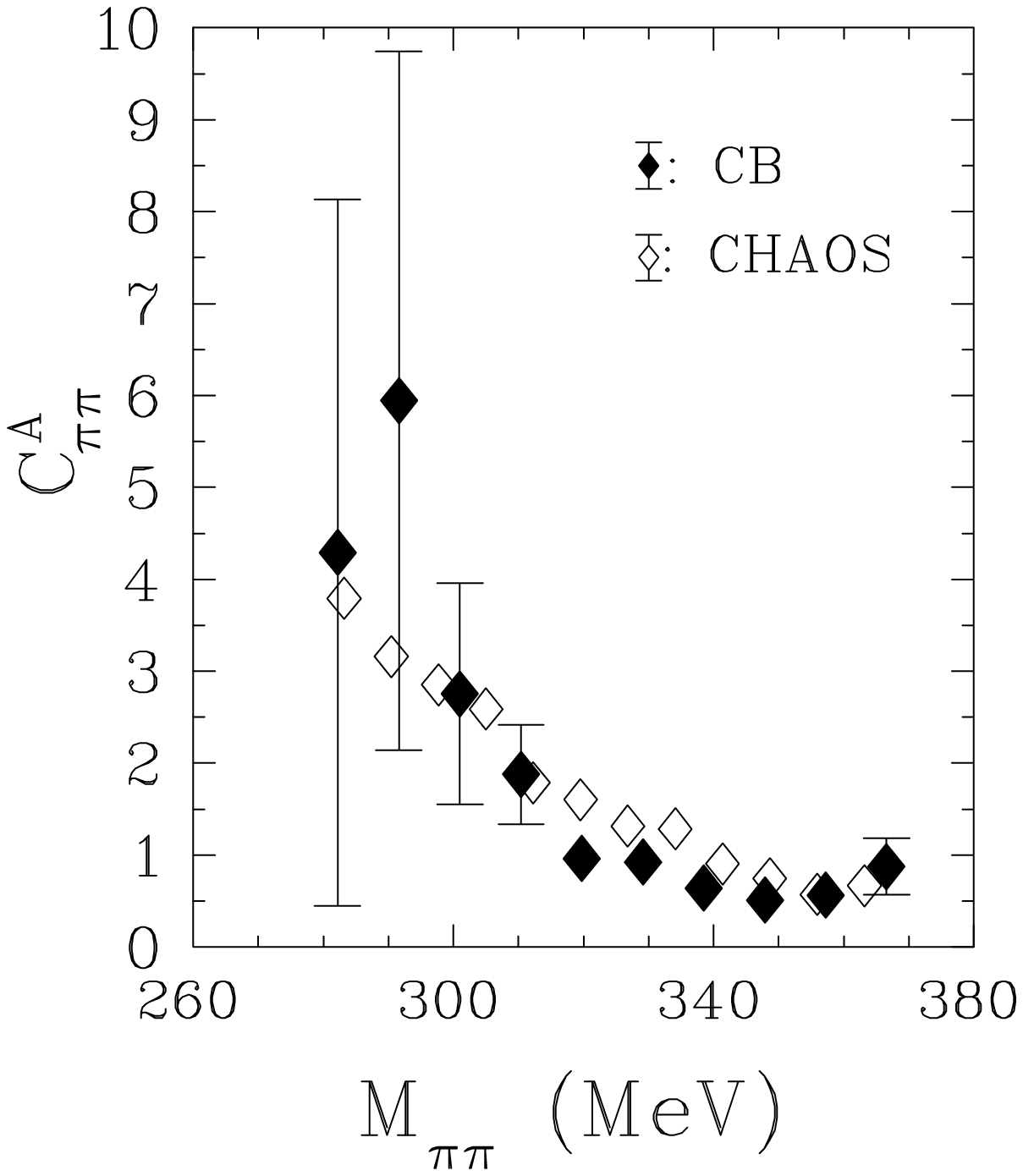}
\caption{The composite ratio ${\cal C}^A_{\pi\pi}$ for $^{12}$C~\cite{Camerini}.}
\label{fig:cratio1}
\end{minipage}
\hspace{\fill}
\begin{minipage}[t]{75mm}
\hspace{-1.0cm}\includegraphics[angle=90,width=20pc]{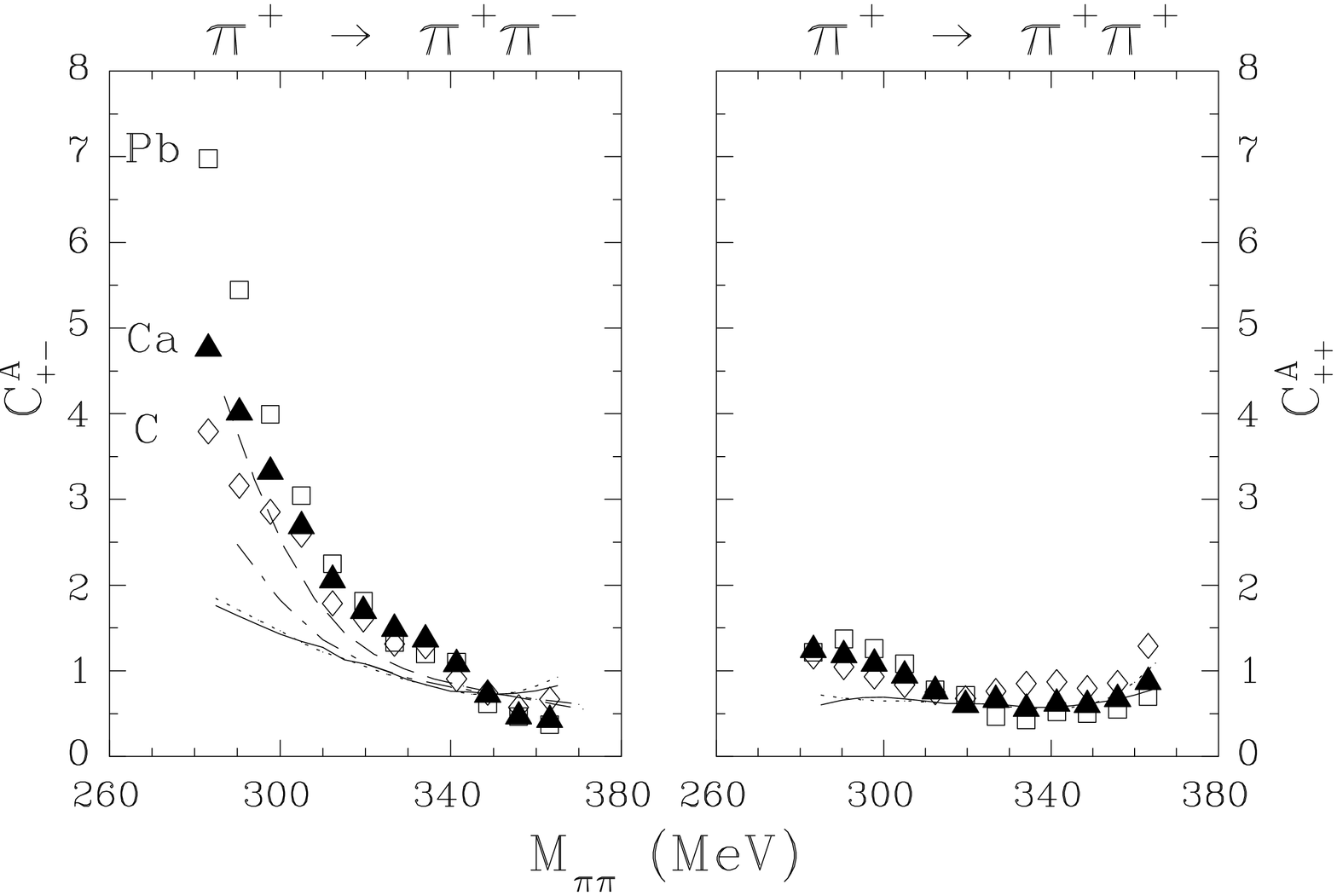}
\caption{The composite ratio $C_{\pi\pi}^A$ for various nuclear targets~\cite{chaos2}.}
\label{fig:cratio2}
\end{minipage}
\end{figure}

What are the experimental signatures? Since the $\sigma$ meson strongly
couples to two-pion states an obvious experiment is the
production of  two $J=I=0$ pions near threshold in nuclei. Such experiments have been
conducted by the CHAOS collaboration at TRIUMF using an incident
$\pi^+$ beam on various nuclear targets~\cite{chaos}
identifying charged pions in the
final state. A second experiment is by the Crystal Ball (CB) collaboration 
at BNL with an incident  
$\pi^-$ beam~\cite{crystalbal}
detecting a $\pi^0$ pair in the final state through coincident $4\gamma$
decay. For the invariant mass distribution of the produced
pion pair both measurements consistently observe 
a significant reshaping as function of mass number as indicated in 
Fig~\ref{fig:cratio1}. Here the composite ratio 
\beq
C_{\pi\pi}^A={\sigma^A(M_{\pi\pi})\over\sigma_T^A}/{\sigma^N(M_{\pi\pi})
\over \sigma_T^{N}}~,
\eeq
is displayed, 
where $\sigma^A(M_{\pi\pi})$ ($\sigma^N(M_{\pi\pi})$) denotes the invariant mass 
distribution
in the nucleus (nucleon), while $\sigma_T^A$ ($\sigma_T^N$) is the corresponding 
total cross section for the $\pi2\pi$ process.
The various curves in the left panel of Fig.~\ref{fig:cratio2} reflect the
theoretical predictions. The full~\cite{vicente}
and dotted~\cite{rdackssw} lines only include contributions from p-wave modifications 
of the pion propagator while the dashed-dotted line~\cite{Hats2} 
only considers the dropping of the $\sigma$ mass without 
dressing the pion loop.
Finally the dashed line~\cite{Schuck2} includes both effects and
gives a reasonable description of the data.

\subsection{The $\rho$ meson and Dileptons}

\begin{figure}[htb]
\bce
\includegraphics[width=35pc]{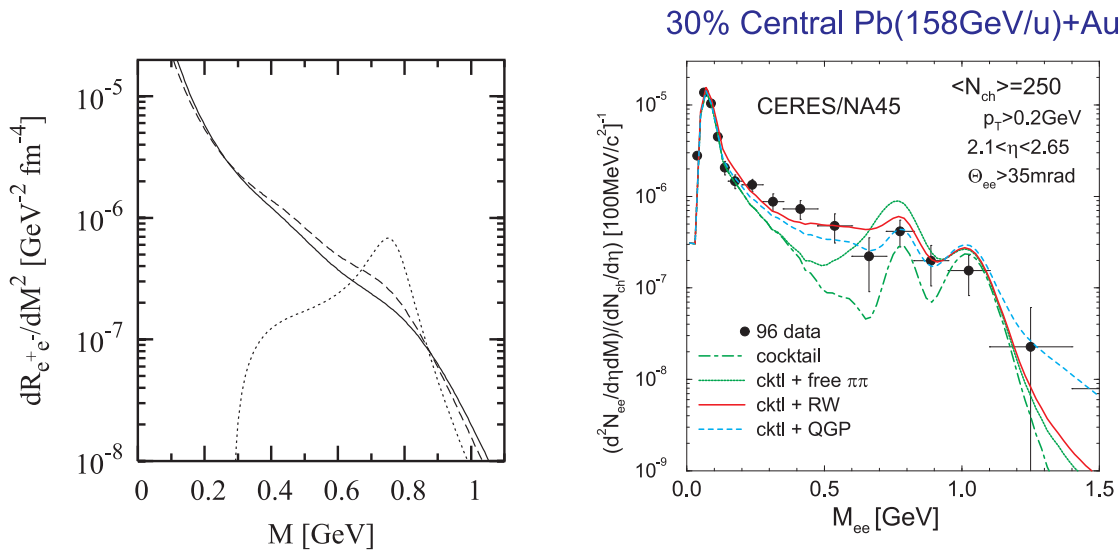}
\caption{Dilepton rates: the left panel displays the theoretical 
prediction at $T$=150 MeV and $\mu_B=3\mu_q$=452 MeV~\cite{Urban1} while the right
panel shows a comparison \cite{RaWa} with the measured rates of the CERES 
collaboration~\cite{CERES}.}
\label{fig:dilept}
\ece
\end{figure}

The $\rho$ meson is featured as a prominent resonance in the $e^+e^-$ annihilation
cross section and therefore plays a central role for dilepton production 
in heavy-ion collisions at invariant masses below 1GeV. In general, 
the production rate is given in terms of the
in-medium electromagnetic current-current correlation function as
\beq
{dR_{l^+l^-}\over d^4q}=f(q_0){L^{\mu\nu}
{\rm Im}{D^{\rm elm}_{\mu\nu}}};\quad
D^{\rm elm}_{\mu\nu}(\omega,\vec q)=-i\!\int\!\!d^4x\,e^{iqx}\theta(x_0)
\tave{[J^{\rm elm}_\mu(x),J^{\rm elm}_\nu(0)]}
\eeq
where $L^{\mu\nu}$ denotes the lepton tensor and $f(q_0)$ is a thermal
Bose factor. Invoking vector dominance, the hadron tensor $D^{\rm elm}_{\mu\nu}$
directly relates to the in-medium properties of the $\rho$ meson and can be
evaluated in the VDM. Two important medium effects can be identified. The first
is the modification of the intermediate two-pion state to which the $\rho$ meson
strongly couples. Here proper care has to be taken to ensure gauge invariance. The
second is a direct coupling to baryonic resonances, most prominently the
$N^*(1550)$ resonance. The inclusion of both effect leads to a large
broadening of the spectral function at high density and temperature and results
in a dramatic reshaping of the dilepton rate as indicated in the left panel of 
Fig.~\ref{fig:dilept}.   
Once the local rate is space-time evolved through a realistic fire ball
expansion until thermal freeze out, and detector acceptances and background
rates from Dalitz decays are properly accounted for, the resulting rates
compare favorably with the measurements of the
CERES collaboration at the CERN SpS~\cite{CERES} (right panel of Fig.~\ref{fig:dilept}).
Also the transverse spectra are well reproduced.

\begin{figure}[htb]
\bce
\includegraphics[width=35pc]{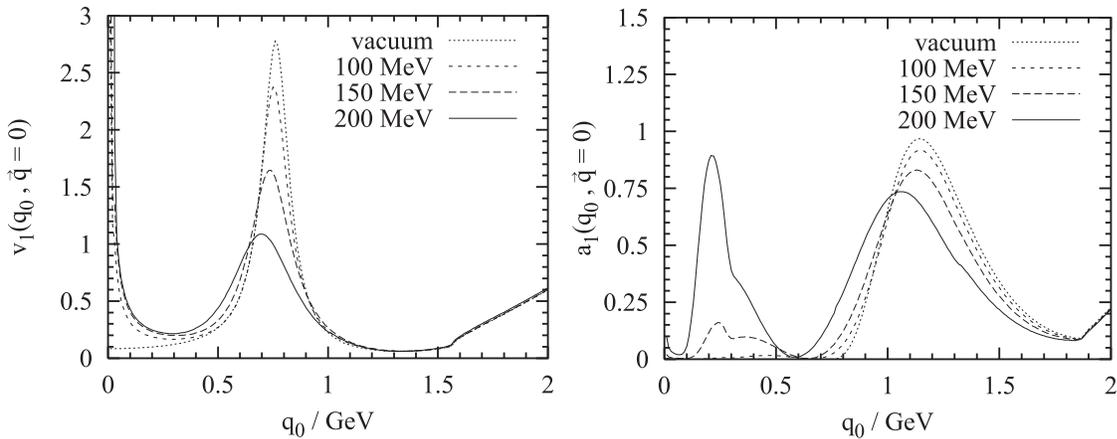}
\caption{The temperature dependence of the vector spectral function (left panel) 
and the axialvector spectral function (right panel) in the chiral 
limit~\cite{Urban3}.}
\label{fig:vacorr}
\ece
\end{figure}

The connection to the restoration of chiral symmetry is not apparent,
however. This requires a simultaneous evaluation of both the vector-
and axialvector correlator. In the vacuum and in the chiral limit 
both are related by two 'Weinberg sum rules'~\cite{Weinb}:
\beq 
\int_0^\infty\!\! ds\,(\rho^\circ_V(s)
-\rho^\circ_A(s))=f_\pi^{2};\qquad
\int_0^\infty\!\! dss\,(\rho^\circ_V(s)
-\rho^\circ_A(s))=0
\eeq
where the first directly links the vacuum spectral functions to $f_\pi$,
the order parameter of chiral symmetry restoration. Similar sum rules also
hold in the hadronic medium~\cite{KaSh} and serve as an
important constraint of models that intend to properly implement chiral
symmetry in the correlators. One such model has recently been proposed
in ref.~\cite{Urban2}. The starting point is the linear sigma model (a non-linear
realization of chiral symmetry could also be chosen). Combining 
the $\rho$- and $a_1$ fields as chiral partners 
\beq
Y^\mu = \vec\rho^\mu\cdot\vec T+\vec a^\mu_{1}\cdot\vec T_5
\eeq
one then writes down the most general Lagrangian consistent with
{\it global} chiral symmetry to a given order in the derivative coupling.
This introduces a set of bare parameters to be determined from the vacuum
phenomenology, in particular the measured vector- and axialvector spectral
functions. In the one-loop approximation one can achieve an excellent
description of the pion electromagnetic form factor, p-wave $\pi\pi$ phase 
shifts, $e^+e^-$ cross sections and $\tau$-decay~\cite{Urban2}. The extension
to finite temperature is straightforward and results in the spectral
distributions displayed in Fig.~\ref{fig:vacorr} for the chiral limit.   
At high temperatures a significant reshaping of the strength distribution is
observed, especially at low energy. Nonetheless the $\rho$ and $a_1$ peak
remain present, even in the vicinity of the phase boundary. The
parameters of the model Lagrangian can be easily adjusted so that, at tree level,
the $\rho$-meson mass is proportional to the chiral condensate 
$\tave{\sigma}$. This is the scenario 
of 'Brown-Rho scaling'~\cite{BR}. To one-loop order,
however, it can be proven analytically that the pole mass of the $\rho$ meson 
remains unchanged to order $T^2$ and only receives contributions of order $T^4$
and higher hence invalidating the 'Brown-Rho scenario'. 

\section{SUMMARY}

Based on general considerations of spontaneous breaking of chiral symmetry 
and its restoration in hot and dense hadronic matter we
have discussed the in-medium properties of $\sigma$- and $\rho$ mesons. The former 
relate to the increased fluctuations of the chiral order parameter which are
accessible through measurements of s-wave isoscalar two-pion correlations in
nuclei. Interesting near-threshold enhancements in the invariant-mass distribution
are observed in $\pi2\pi$ reactions, which can be interpreted as a signal of
the partial restoration of chiral symmetry. The $\rho$ meson and its medium
modification plays an important role in the dilepton production in relativistic
heavy-ion collisions. Realistic calculations of the spectral function in a 
hot and dense fireball indicate a significant broadening due to collisions with 
baryons and mesons and explain the observed production rates. The 
relation to chiral symmetry is not obvious, however. To address this issue 
one has to consider the vector- and
axialvector correlators simultaneously. Invoking global chiral symmetry in
constructing a chiral Lagrangian from elementary $\sigma,\pi,\rho$ and $a_1$
fields it can be shown that, in the chiral limit, there is no direct relationship
between the pole mass of the $\rho$ meson and the chiral condensate.


\begin{thebibliography}{9}
\bibitem{Nege} J. W. Negele, Nucl. Phys. Proc. Suppl. 73 (1999) 92.
\bibitem{SchaSh} T. Sch\"afer, E. V. Shuryak, Rev. Mod. Phys. 70 (1998) 323.
\bibitem{Nambu} Y. Nambu and G. Jona-Lasinio, Phys. Rev. 122 (1961) 345.
\bibitem{OeBuWa} M. Oertel, M. Buballa, J. Wambach, Nucl. Phys. A676 (2000) 247;\\
         M. Oertel, M. Buballa, J. Wambach, hep-ph/0008131.
\bibitem{Saku} J. J. Sakurai, Ann. Phys. 11 (1960) 1. 
\bibitem{Aoui1} Z. Aouissat, P. Schuck, J. Wambach, Nucl. Phys. A618 (1997) 402.  
\bibitem{Karsch1} F. Karsch, hep-ph/0103314.
\bibitem{Karsch2} F. Karsch, Nucl. Phys. A590 (1995) 367.
\bibitem{PiWi}R. Pisarski, F. Wilczek, Phys. Rev. D29 (1984) 338.
\bibitem{Schuck2}Z. Aouissat, G. Chanfray, P. Schuck, J. Wambach, Phys. Rev. C61 
(2000) 12202.
\bibitem{Hats1} D. Jido, T. Hatsuda, T. Kunihiro, hep-ph/0008076. 
\bibitem{chaos} F. Bonutti et al., Phys. Rev. Lett. 77 (1996) 603.  
\bibitem{crystalbal} A. B. Starostin et al., Phys. Rev. Lett. 85 (2000) 5539.
\bibitem{Camerini} P. Camerini et al., submitted to Phys. Rev. C. 
\bibitem{chaos2} F. Bonutti et al., Nucl. Phys. A677 (2000) 213. 
\bibitem{vicente} M. J. Vicente Vacas, E. Oset,  Phys. Rev. C60 (1999) 64621.
\bibitem{rdackssw} R. Rapp et al., Phys. Rev. C59 (1999) R1237.
\bibitem{Hats2} T. Hatsuda, T. Kunihiro, H. Shimizu, Phys. Rev. Lett. 82 (1999) 2840.
\bibitem{Urban1} M. Urban, M. Buballa, J. Wambach, Nucl. Phys. A673 (2000) 357.
\bibitem{RaWa} R. Rapp, J. Wambach, Adv. Nucl. Phys. 25 (2000) 1.
\bibitem{CERES} G. Agakichiev et al., Phys. Rev. Lett. 75 (1995) 1272.
\bibitem{Weinb} S. Weinberg, Phys. Rev. Lett. 18 (1967) 507.
\bibitem{KaSh} J. Kapusta, E. V. Shuryak, Phys. Rev. D49 (1994) 4694.
\bibitem{Urban2} M. Urban, M. Buballa, J. Wambach, hep-ph/0102260.
\bibitem{Urban3} M. Urban, M. Buballa, J. Wambach, work in progress.
\bibitem{BR} G. E. Brown, M. Rho, Phys. Rev. Lett. 66 (1991) 2720.
\end{thebibliography}
\end{document}